\documentclass[conference]{IEEEtran}
\IEEEoverridecommandlockouts
\usepackage{cite}
\usepackage{url} 
\usepackage{amsmath,amssymb,amsfonts}
\usepackage{graphicx}
\usepackage{textcomp}
\usepackage{xcolor}
\usepackage{algorithm}
\usepackage{algpseudocode}
\usepackage{mathrsfs}
\usepackage{subcaption}
\usepackage[compact]{titlesec}
    \titlespacing{\section}{0pt}{0.5ex}{0.4ex}
    \titlespacing{\subsection}{0pt}{0.4ex}{0.4ex}
    \titlespacing{\subsubsection}{0pt}{0.5ex}{0.4ex}

\usepackage[
  top=0.75in,
  bottom=1.3in,
  left=0.75in,
  right=0.75in
]{geometry}

\usepackage[caption=false,font=footnotesize]{subfig}

\def\BibTeX{{\rm B\kern-.05em{\sc i\kern-.025em b}\kern-.08em
    T\kern-.1667em\lower.7ex\hbox{E}\kern-.125emX}}
\begin{document}

\title{
Jamming-Resilient PRB Reservation for Latency-Critical O-RAN Network Slicing \vspace{-0.1in}
}

\author{Elahe Delavari, and Junaid~Farooq\\
$^\dagger$Department of Electrical and Computer Engineering,
University of Michigan-Dearborn,\\ Dearborn, MI, 48128 USA, Emails: \{elahed, mjfarooq\}@umich.edu. \vspace{-0.1in}
}

\maketitle

\begin{abstract}

Open radio access network (O-RAN) architectures enable near real-time, software-driven control of network slicing through programmable xApps deployed on the near-real-time RAN Intelligent Controller (near-RT RIC). In industrial 5G downlink systems, adversarial jamming can abruptly reduce the effective physical resource block (PRB) capacity, triggering queue buildup and persistent latency violations, particularly in the presence of low spectral efficiency cell edge user equipments. This paper proposes a reserve-based resilience framework for PRB allocation in sliced O-RAN deployments. A finite pool of reserved PRBs is controlled by a near-RT RIC xApp that provides hybrid mitigation by proactively clearing backlog to build latency margin and reactively allocating reserve capacity during jammer active intervals. We formulate reserve activation as a constrained sequential decision problem and design a masked Deep Q-Network to learn effective control policies under non-stationary jamming. Simulation results show substantial reductions in URLLC latency violations and improved reserve efficiency compared to reactive baselines.
\end{abstract}


\section{Introduction}
\label{sec:Introduction}

Industrial wireless networks must simultaneously support heterogeneous services ranging from high-throughput sensing and video analytics to ultra-low latency and highly reliable control for robotics and automation \cite{luo2023field}. In 5G and beyond systems, these requirements are commonly realized through network slicing, where enhanced mobile broadband (eMBB) and ultra-reliable low-latency communication (URLLC) services share a common radio access infrastructure. In industrial settings, disruptions to slice performance can directly impact control stability, safety, and productivity, making resilience a first order design requirement.

The open radio access network (O-RAN) architecture enables software driven and near-real-time control of radio resource management through the near-real-time RAN intelligent controller (Near-RT RIC) \cite{polese2023understanding,balasubramanian2021ric}. By supporting operator deployed xApps and standardized control interfaces, O-RAN allows slice level resource allocation to adapt to time varying network conditions. While this flexibility enables advanced control strategies, it also requires that slicing mechanisms remain robust under abrupt physical layer capacity degradation caused by adversarial jamming or strong interference.

\begin{figure}[t]
    \centering
    \includegraphics[width=0.8\linewidth]{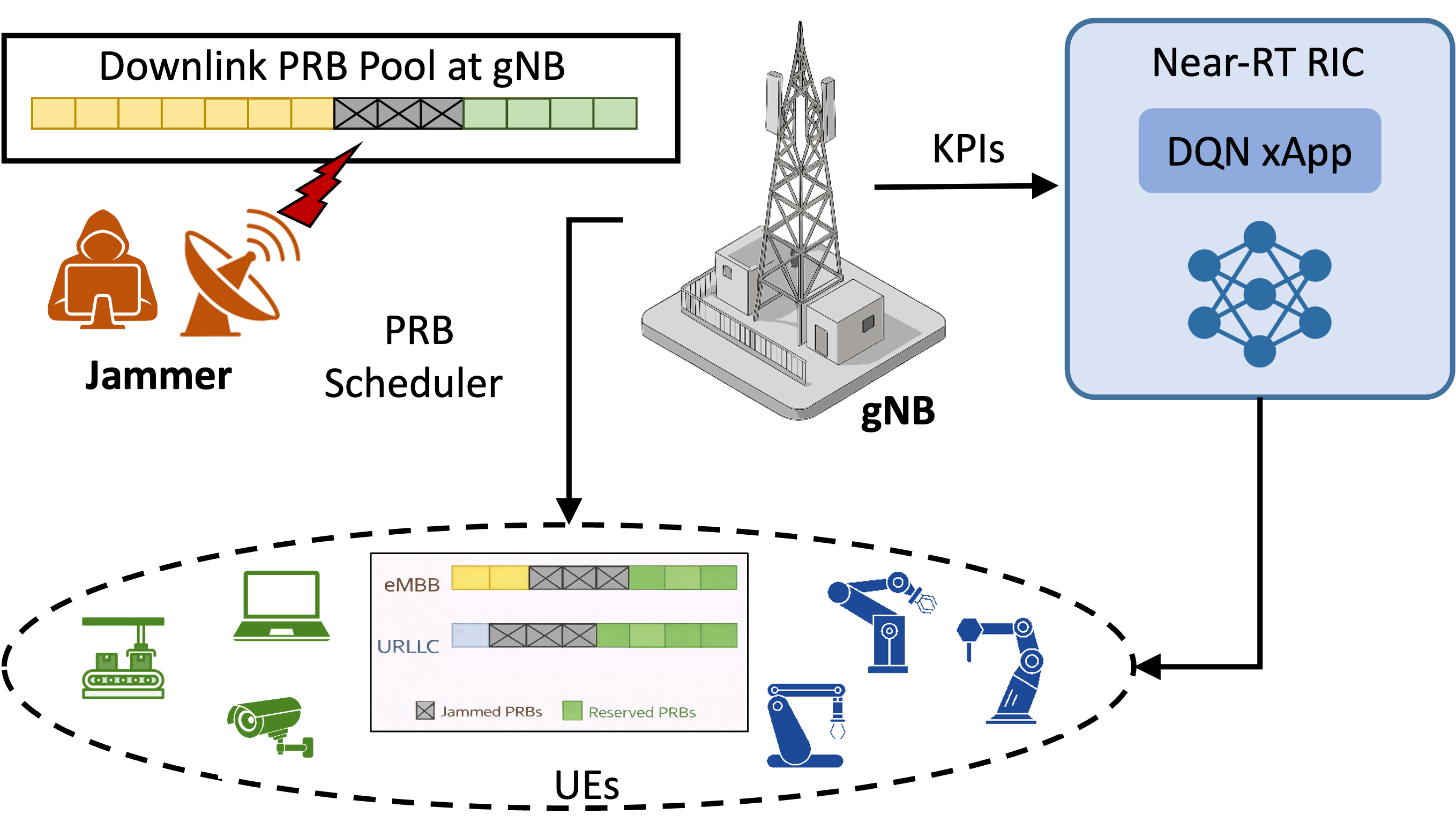}
    \caption{System overview of the resilient PRB allocator. The Near-RT RIC xApp adjusts PRB quotas from a reserved pool to mitigate jamming-induced performance losses.}
    \label{fig:system_overview}
\end{figure}
Jamming attacks reduce the effective number of schedulable physical resource blocks (PRBs), leading to immediate capacity loss. An important and often overlooked factor is the interaction between such capacity shocks and queueing dynamics driven by spatial heterogeneity among user equipments. In industrial cells, a subset of user equipments (UEs) typically operates near the cell edge with low spectral efficiency and consumes a disproportionate share of PRBs. As a result, backlog may accumulate even under nominal conditions. When jamming occurs, even briefly, the resulting reduction in effective PRB capacity can trigger persistent URLLC latency inflation that outlasts the jammer active interval.

Existing jamming mitigation approaches in O-RAN primarily rely on reactive control. Recent work integrates KPI monitoring, jamming detection, and slice based PRB reallocation through Near-RT RIC mitigation pipelines \cite{moore2025demonstrating}. Learning assisted anti jamming techniques using deep reinforcement learning and federated learning have also been proposed \cite{thanh_anti-jamming_2022,sharma2022mitigating,asemian_anti-jamming_2025,houda_federated_2024}. While effective during jammer active periods, these approaches implicitly assume that the system enters the disruption in a relatively un-congested state. In backlog prone regimes dominated by low spectral efficiency UEs, purely reactive mitigation may be insufficient once queue buildup has already occurred.

Resource reservation has been widely studied as a mechanism to hedge uncertainty in wireless systems, including dynamic spectrum reservation in cognitive radio networks \cite{balapuwaduge2018dynamic,el2020centralized}, hybrid reservation for heterogeneous users \cite{abbas2020spectrum,khan2021enhanced}, and reservation based medium access control \cite{luis2016novel}. However, reservation has not been explicitly explored as a resilience mechanism for mitigating adversarial capacity shocks in O-RAN based network slicing. Motivated by this gap, this paper introduces a reserve-based resilience framework for PRB allocation in sliced O-RAN deployments. A finite pool of reserved PRBs is controlled by a Near-RT RIC xApp to provide both proactive and reactive mitigation under jamming, as illustrated in Fig.~\ref{fig:system_overview}. The main contributions of this work are summarized as follows:
\begin{itemize}
    \item We formulate jamming resilient PRB allocation in sliced O-RAN systems as a sequential decision problem that accounts for backlog driven latency amplification and finite reserve budgets.
    \item We propose a reserve-based hybrid mitigation strategy that combines proactive backlog clearance with reactive reserve allocation during jammer-active intervals.
    \item We design a masked Deep Q-Network xApp for the Near-RT RIC that learns feasible reserve activation policies under non-stationary jamming.
    \item Simulation results demonstrate significant reductions in URLLC latency violations and improved reserve efficiency compared to reactive baselines.
\end{itemize}


\section{System Model}
\label{sec:system_model}

\subsection{Network Setup}
We consider a single-cell 5G downlink system with one gNodeB (gNB) serving a macro cell of radius $R_{\text{cell}}$ over bandwidth $W$ at carrier frequency $f_c$.
The gNB transmits with power $P_{\text{tx}}$.
Let $\mathcal{U}=\{1,\ldots,K\}$ denote the set of UEs served by the gNB, where $K \triangleq |\mathcal{U}|$.
The UE set is partitioned into two slice-specific subsets:
$\mathcal{U}=\mathcal{U}_e \cup \mathcal{U}_u$, with $\mathcal{U}_e \cap \mathcal{U}_u=\emptyset$,
where $\mathcal{U}_e$ and $\mathcal{U}_u$ denote the eMBB and URLLC UE sets, respectively.
We denote the number of UEs per slice by $K_e \triangleq |\mathcal{U}_e|$ and $K_u \triangleq |\mathcal{U}_u|$, hence $K=K_e+K_u$.
At the beginning of each episode, UEs are placed uniformly at random within a disk of radius $R_{\text{spawn}}$ centered at the gNB and remain fixed during the episode.

The system bandwidth is partitioned into $N_{\text{total}}$ PRBs, which form the basic time--frequency scheduling unit in 5G. Let $\eta_{\text{DL}} \in (0,1]$ denote the downlink fraction. The nominal downlink PRB capacity is
$
C_{\max} \triangleq \eta_{\text{DL}} \, N_{\text{total}} .
$
Two downlink slices coexist: eMBB ($s=e$) and URLLC ($s=u$). 
Each slice has a configured minimum PRB quota $q_s^{\min}$.
In addition, a finite reserved resiliency pool of size $Z$ PRBs can be activated to mitigate performance degradation.
Let $n_s(t)\in \mathbb{Z}_{\ge 0}$ denote the number of reserved PRBs allocated to slice $s$ at decision step $t$.
The resulting slice quota is
$
q_s(t) = q_s^{\min}(t) + n_s(t), \quad s\in\{e,u\}.
$
we should note that the $q_s^{\min}(t)$ is changing based on the jamming as the number of PRBs are decreased.
Given the slice quota $q_s(t)$, the gNB allocates PRBs among UEs in slice $s$ proportionally to each UE's instantaneous demand. The resulting per-UE allocation $\mathcal{V}_k(\tau)$ is then used to compute the achievable service rate.

\subsection{Channel Model and Achievable Data Rate}
\label{subsec:channel}

Let $\mathbf{p}_g\in\mathbb{R}^2$ and $\mathbf{p}_i\in\mathbb{R}^2$ denote the positions of the gNB and UE $i$, respectively. The distance between the gNB and UE $i$ is
$
d_i \triangleq \lVert \mathbf{p}_i - \mathbf{p}_g \rVert_2 .
$
Large-scale path loss follows the 3GPP urban macro (UMa) non-line-of-sight (NLOS) model~\cite{ETSI_TR_138_901}. 
The LOS and NLOS path losses in dB are given by
\begin{equation}
\begin{aligned}
PL^{\text{LOS}}(d_i) 
&= 28 + 22\log_{10}(d_i) + 20\log_{10}(f_c), \\
PL^{\text{NLOS}}(d_i) 
&= 13.54 + 39.08\log_{10}(d_i) + 20\log_{10}(f_c) \\
&\quad - 0.6\big(h_{\text{UE}} - 1.5\big).
\end{aligned}
\label{eq:pl_los_nlos}
\end{equation}
where $h_{\text{UE}}$ denotes the UE antenna height (meters). The effective path-loss is
 $
 PL(d_i) = \max\!\big(PL^{\text{LOS}}(d_i),\, PL^{\text{NLOS}}(d_i)\big).
 $
The received downlink power at UE $i$ is
$
P_{\text{rx},i}^{\text{(dBm)}} = P_{\text{tx}}^{\text{(dBm)}} - PL(d_i),
$
which is converted to linear scale (Watts). Thermal noise power is
$
N = k_B \, T_{\text{UE}} \, B,
$
where $k_B$ is Boltzmann’s constant, $T_{\text{UE}}$ is the UE noise temperature, and $B$ is the system bandwidth.
Assuming a single serving gNB and neglecting inter-cell interference, the downlink signal-to-noise ratio (SNR) is
\begin{equation}
\mathrm{SNR}_i = \frac{P_{\text{rx},i}^{\text{(W)}}}{N},
\label{eq:snr_linear}
\end{equation}


The SNR expressed in dB is mapped to a channel quality indicator (CQI) following 3GPP TS 38.214 \cite{3gpp_38214}. This method utilizes a thresholding function $\mathscr{T}(\cdot)$ such that $\text{CQI}_i = \mathscr{T}(\text{SNR}_i|_{\text{dB}})$.
The CQI value is subsequently used to select the Modulation and Coding Scheme (MCS) using lookup tables derived from 3GPP TS 38.214 Table 5.1.3.1-2, yielding a modulation order $M_i$ and a target code rate $c_i \in [0, 1024]$.
Let $\mathcal{V}_i(\tau)$ denote the number of PRBs allocated to UE $i$ during slot $\tau$.
The achievable downlink rate (bits/s) is computed as
\begin{equation}
r_i(\tau) = \frac{\mathcal{V}_i(\tau)\, \mathcal{N}_{\text{RE}} \, M_i \, \big(c_i/1024\big)}{\mathcal{T}_{\text{slot}}},
\label{eq:rate_prb_slot}
\end{equation}
where $\mathcal{N}_{\text{RE}}$ is the number of resource elements per PRB per slot and $\mathcal{T}_{\text{slot}}$ is the slot duration. The slice throughput $\mathscr{R}_s(t)$ denotes the total downlink data rate achieved by
slice $s$ at time $t$, computed as 
$\mathscr{R}_s(t) = \sum_{i \in u_s} r_i(t)$.

\subsection{Traffic and Queueing Dynamics}
\label{subsec:latency}
Downlink traffic is buffered at the gNB on a per-UE basis.
For each UE $i$, the gNB maintains a first-in-first-out (FIFO) queue of chunks $\{(a_{i,j}, b_{i,j})\}_j$, where $a_{i,j}$ is the arrival time and $b_{i,j}$ is the number of bytes in chunk $j$.
Traffic is replayed from per-UE DL traces, i.e., whenever a trace bin becomes due, the corresponding bytes are enqueued with timestamp equal to the current simulator time.
Over a scheduling interval of duration $\Delta T_{\text{step}}$,
let $r_i(t)$ denote the effective data rate (bits/s) available to UE $i$.
The corresponding service capacity in bytes is
$
S_i(t) = \frac{r_i(t)\,\Delta T_{\text{step}}}{8}.
$
\label{eq:service_bytes}
Let $X_i(t)$ be the number of bytes actually served for UE $i$ during this interval, obtained by draining up to $S_i(t)$ bytes from the FIFO queue.
If $X_i(t)>0$, the reported downlink latency equals the bytes-weighted queueing delay of the served bytes:
\begin{equation}
\mathcal{L}_i(t) \triangleq 
\frac{\sum_{j\in\mathcal{D}_i(t)} b_{i,j}\,\big(t-a_{i,j}\big)}
{\sum_{j\in\mathcal{D}_i(t)} b_{i,j}},
\label{eq:weighted_delay}
\end{equation}
where $\mathcal{D}_i(t)$ is the set of FIFO chunks dequeued during the interval.
If $X_i(t)=0$ and the FIFO queue is non-empty, the reported latency is the head-of-line waiting time 
$
\mathcal{L}_i(t) \triangleq t - a_{i,\text{HOL}},
$
where $a_{i,\text{HOL}}$ is the arrival time of the oldest queued bytes.
For slice $s$ at time $t$, the simulator reports the mean of per‑UE downlink latencies as
$
L_s(t) \triangleq \frac{1}{|\mathcal{U}_s(t)|} \sum_{i \in \mathcal{U}_s(t)} \mathcal{L}_i(t).
$

\subsection{Attack Model}
\label{sec:jamming_model}

We consider active jamming attacks that degrade the downlink capacity of the radio access network by reducing the number of PRBs effectively available for scheduling.
Rather than explicitly modeling jammer waveforms or physical-layer interference, we adopt a scheduler-level abstraction in which jamming manifests as a time-varying reduction in the usable downlink PRB budget.
Let $J(t)\ge 0$ denote the jamming severity at controller decision step $t$, expressed as the number of PRBs rendered unavailable due to the attack.
The effective downlink PRB budget is
$
C_{\mathrm{eff}}(t) = C_{\max} - Z - J(t),
$
where $C_{\max}$ is the nominal downlink PRB capacity of the cell.
To mitigate such capacity loss, the controller may activate PRBs from a finite resilience pool with total budget $Z$ over the horizon $T$, subject to a per-step activation cap $N_{\text{pool}}$.
We assume $J(t)$ takes values from a finite discrete set $\mathcal{J}$. 

\section{Resilient PRB Reservation Framework}
\label{sec:methodology}

This section presents a control-theoretic formulation of jamming-resilient resource allocation in O-RAN network slicing and motivates the use of learning-based control. We focus on the design and simulation-based evaluation of the reserve-control logic intended for Near-RT RIC xApps, rather than a full O-RAN-compliant software implementation. We first define an idealized stochastic optimization problem that captures the tradeoff between service protection and finite reserve usage under jamming. We then show why this formulation is intractable in practice, leading to the design of a DRL based xApp for near-real-time reserve control.


\subsection{Problem Formulation}
We define resilience as the ability of the slicing controller to limit service-level agreement (SLA) violations over a finite horizon $T$ under stochastic jamming. At each controller decision step $t$, the Near-RT RIC selects the number of reserved PRBs $n_s(t)$ allocated to each slice $s\in\{e,u\}$, which in turn determines slice quotas, per-UE scheduling, queue evolution, and achieved performance metrics as described in Section~II.

Let $L_u(t)$ denote the resulting mean URLLC downlink latency at time $t$ and let $\mathscr{R}_e(t)$ denote the resulting aggregate eMBB downlink throughput. These quantities are induced by the system dynamics and depend on the reserve allocation decisions $\{n_u(\tau),n_e(\tau)\}_{\tau\le t}$, the jamming process $J(t)$, channel conditions, and traffic arrivals. Importantly, $L_u(t)$ and $\mathscr{R}_e(t)$ do not admit closed-form expressions due to nonlinear queueing dynamics and adaptive scheduling.

To quantify SLA violations, we introduce non-negative slack variables $\delta_u(t)$ and $\delta_e(t)$ capturing URLLC latency excess and eMBB throughput shortfall, respectively. The resilience-aware reserve allocation problem is formulated as the following constrained stochastic optimization:


\begin{equation}
\min_{\{n_u(t), n_e(t)\}} \mathbb{E} \left[ \sum_{t=1}^{T} \left( \omega_u \delta_u(t) + \omega_e \delta_e(t) + \omega_p \left( n_u(t) + n_e(t) \right) \right) \right]
\label{eq:objective}
\end{equation}
\text{Subject to:}
\begin{align}
    & L_u(n_u(t), J(t)) \le L_{\text{target}} + \delta_u(t) & \forall t \in \{1, \dots, T\} \label{eq:const_lat} \\
    & \mathscr{R}_e(n_e(t), J(t)) \ge \mathscr{R}_{\text{target}} - \delta_e(t) & \forall t \in \{1, \dots, T\} \label{eq:const_thr} \\
    & n_u(t) + n_e(t) \le N_{\text{pool}} & \forall t \in \{1, \dots, T\} \label{eq:const_pool} \\
    & \sum_{t=1}^{T} \left( n_u(t) + n_e(t) \right) \le Z \label{eq:const_budget} \\
    & n_u(t), n_e(t) \in \mathbb{Z}^+, \quad \delta_u(t), \delta_e(t) \ge 0 \label{eq:const_nonneg}
\end{align}
The expectation is taken over exogenous uncertainties including UE locations, channel realizations, traffic arrivals, and the jamming process $J(t)$. Constraint~\eqref{eq:const_budget} introduces explicit time coupling across decision steps. Excessive reserve activation early in the horizon reduces the controller’s ability to mitigate future disruptions, while insufficient activation allows backlog accumulation that can cause persistent URLLC latency inflation even after jamming subsides.


\subsection{DRL-Based Reserve Allocation}

The optimization problem in Section~III-A characterizes the desired resilience behavior but cannot be solved directly in near-real-time. The key difficulty lies in the fact that the performance metrics $L_u(t)$ and $\mathscr{R}_e(t)$ are emergent quantities induced by queueing dynamics, adaptive scheduling, channel variations, and stochastic jamming. These dynamics are nonlinear, history dependent, and do not admit tractable closed-form models. Moreover, the jamming process $J(t)$ is unknown and non-stationary, making future capacity degradation inherently unpredictable. As a result, conventional model-based optimization is infeasible for Near-RT RIC operation.

To address these challenges, we formulate reserve activation as a Markov Decision Process (MDP) and learn a control policy directly from interaction with the environment. The controller operates at discrete decision steps indexed by $t$, observes a slice-level system state, and selects reserve allocation actions that influence subsequent queue evolution and performance. This learning-based approach enables adaptive control under uncertainty while implicitly accounting for delayed effects of past decisions.

Formally, the reserve control problem is modeled as an MDP
$\langle \mathcal{S}, \mathcal{A}, \mathcal{P}, \mathcal{R}, \gamma \rangle$,
where $\mathcal{S}$ denotes the observable system state, $\mathcal{A}$ the action space, $\mathcal{P}$ the unknown transition dynamics induced by traffic, channels, scheduling, and jamming, $\mathcal{R}$ the reward function, and $\gamma\in[0,1]$ the discount factor. A learning-based xApp deployed on the Near-RT RIC seeks to learn a policy $\pi:\mathcal{S}\rightarrow\mathcal{A}$ that minimizes cumulative SLA violations and reserve usage, consistent with the objective in~\eqref{eq:objective}.



\subsubsection{State Representation}
The state $s(t)\in\mathbb{R}^{15}$ provides a normalized view of demand, allocation,
performance, and jamming:
\begin{equation}
\begin{aligned}
s(t)=\big[
\tilde{D}_e,\tilde{D}_u,
\tilde{q}_e,\tilde{q}_u,
\tilde{q}^{min}_e,\tilde{q}^{min}_u,
\bar{Q}_e,\bar{Q}_u,
\tilde{\mathscr{R}}_e,\tilde{\mathscr{R}}_u,\\
\tilde{L}_u,
\tilde{C}_{eff},
\tilde{N}_{used},
\tilde{a}_{e,\mathrm{prev}},\tilde{a}_{u,\mathrm{prev}}
\big].
\end{aligned}
\end{equation}
Here, $\tilde{D}_s$ is normalized PRB demand, $\tilde{q}_s$ and $\tilde{q}^{min}_s$ are current/base PRB quotas,
$\bar{Q}_s$ is normalized backlog, $\tilde{\mathscr{R}}_s$ is normalized throughput, $\tilde{L}_u$ is normalized URLLC latency,
$\tilde{C}_{eff}$ is the effective PRB ratio, $\tilde{N}_{used}$ is the fraction of reserved PRBs in
use, and $\tilde{a}_{s,\mathrm{prev}}$ is the previous action for slice $s\in\{e,u\}$.

\subsubsection{Action Space}
We use a discrete joint action $a(t)=(\Delta n_e(t), \Delta n_u(t))$ that adjusts the reserved PRBs assigned to each slice.
Each component is selected from a finite step set $\mathcal{D}\subset\mathbb{Z}$:
\begin{equation}
\Delta n_s(t)\in \mathcal{D}, \quad s\in\{e,u\}, 
\qquad \mathcal{A}=\mathcal{D}\times \mathcal{D}.
\end{equation}
The reserve update is
$
n_s(t)=n_s(t-1)+\Delta n_s(t).
$
To enforce feasibility and speed up learning, we apply action masking~\cite{kanervisto2020action} and restrict to
$\mathcal{A}_{\mathrm{valid}}(s(t))\subseteq\mathcal{A}$ by discarding actions that violate slice quotas,
exceed the effective capacity budget, or returns more reserved PRBs than currently borrowed.

\subsubsection{Reward Design}
The reward function is constructed as a smooth surrogate of the stochastic optimization objective in~\eqref{eq:objective}. At each decision step, the controller receives
\begin{equation}
\mathcal{R}(t)= -\big(P_e(t)+P_u(t)+P_{\mathrm{res}}(t)\big).
\end{equation}
Define the eMBB throughput gap and URLLC latency excess as
\begin{equation}
G_e(t)=[\mathscr{R}_{\mathrm{target}}-\mathscr{R}_e(t)]^+,\qquad
E_u(t)=[L_u(t)-L_{\mathrm{target}}]^+,
\end{equation}
and apply log-scaled QoS penalties
\begin{equation}
P_e(t)=\omega_e\ln\!\big(1+G_e(t)\big),\qquad
P_u(t)=\omega_u\ln\!\big(1+E_u(t)\big).
\end{equation}
To encourage recovery under jamming while avoiding unnecessary reserve activation, we waive the reserve penalty during
violations:
\begin{equation}
P_{\mathrm{res}}(t)=
\begin{cases}
0, & G_e(t)>1 \ \text{or}\  E_u(t)>1,\\[2pt]
\omega_p\,\dfrac{n_e(t)+n_u(t)}{Z}, & \text{otherwise},
\end{cases}
\end{equation}

\subsection{DQN Agent}
\label{subsec:dqn_design}
We model reserved-PRB control as an MDP and learn a discrete policy with DQN. The xApp observes a 15-D normalized state
and selects a joint action $a(t)=(\Delta n_e,\Delta n_u)$. Operational constraints are enforced by action masking; both
$\epsilon$-greedy selection and the TD target maximize only over $\mathcal{A}_{\mathrm{valid}}(\cdot)$.

\subsubsection{Q-network and training}
We parameterize $Q(s,a;\theta)$ by an MLP with two hideen layers of 256 followed by ReLU unites and an output layer of size $|\mathcal{A}|$. We train this network train with replay $10^5$, batch 64 and a target network updated every $2000$ steps. For $(s,a,r,s')$,
\begin{equation}
y=r+\gamma\max_{a'\in\mathcal{A}_{\mathrm{valid}}(s')}Q(s',a';\theta^-),\qquad \gamma=0.99,
\end{equation}
and we minimize $\mathbb{E}\!\left[(y-Q(s,a;\theta))^2\right]$ using Adam ($10^{-4}$). Exploration is $\epsilon$-greedy
over $\mathcal{A}_{\mathrm{valid}}(s)$ with $\epsilon$ annealed linearly from 1.0 to 0.05. 


\section{Simulation Results}
\subsection{Simulation setup}
\begin{figure}[t]
    \centering
    \includegraphics[width=0.40\textwidth]{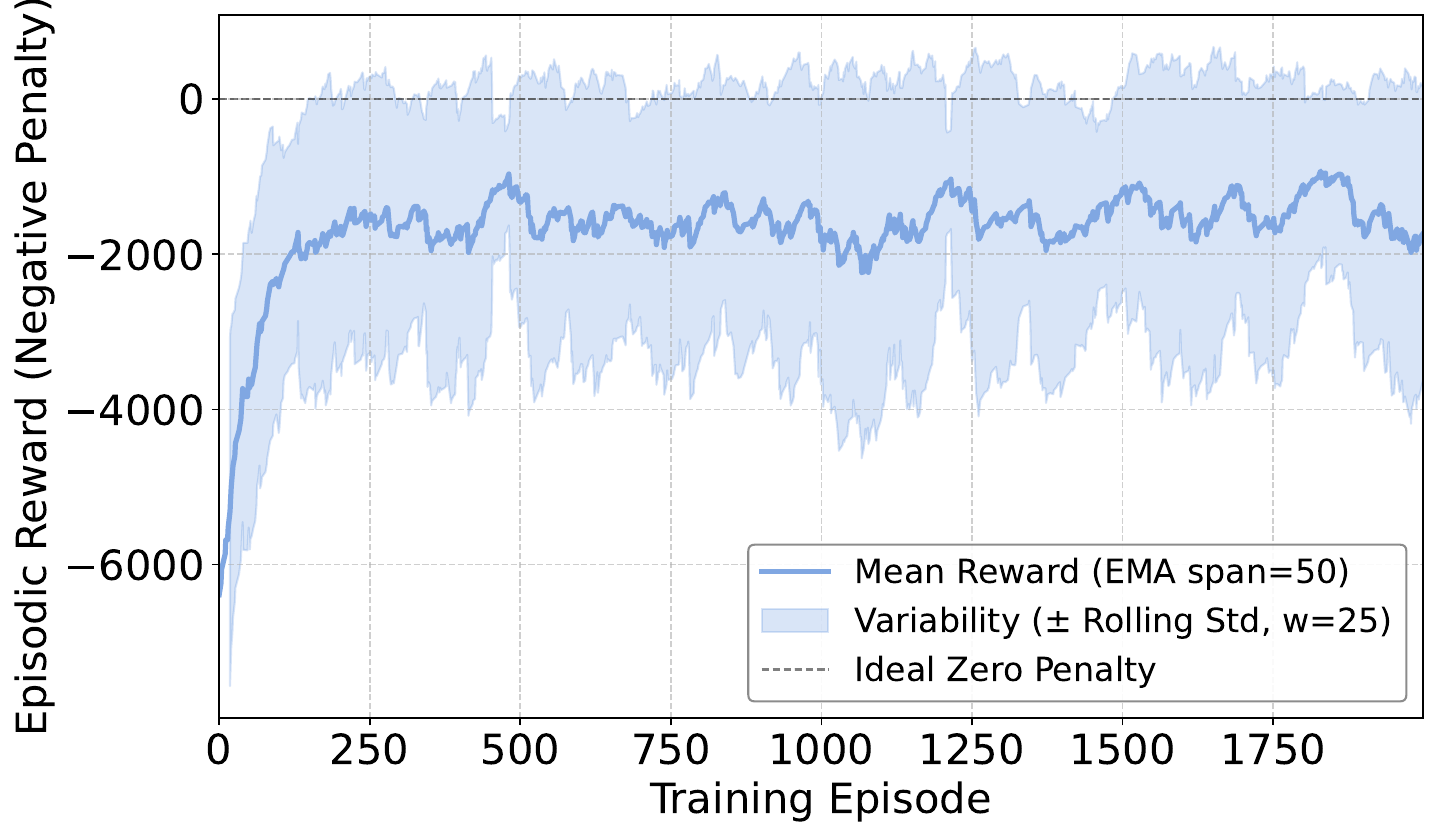}
    \caption{Training convergence of the DQN-based xApp. \vspace{-0.1in}}
    \label{fig:convergence}
\end{figure}

We evaluate our xApp in a Python-based AI-RAN Simulator~\cite{airansim}. Training uses $2{,}000$ episodes of $10$\,s each with
$1$\,ms simulator steps and a control period of $10$\,ms. Fig.~\ref{fig:convergence} shows the episodic training return. 
We adopt a periodic on–off jamming model, commonly used in the wireless literature to capture time-varying adversarial interference \cite{pirayesh_jamming_2022}. During jammer-ON intervals, the effective PRB budget is reduced according to $J(t)$; during jammer-OFF intervals, $J(t)=0$ and the system operates at full capacity. To evaluate resilience across attack intensities, we consider a discrete set of severity levels $\mathcal{J}$ (Table~\ref{tab:sim_params}).
Within each jammer-ON interval, the severity follows one of four non-stationary profiles: fixed—constant within the interval; increasing—monotonically increasing; decreasing—monotonically decreasing; or random—time-varying under a randomized process.

The jammer follows a $50\%$ duty
cycle with period $2500$ simulator steps, yielding four jammer-ON intervals per episode.
Testing uses $50$ episodes with identical UE-placement seeds across methods for fair comparison.
We compare four policies, namely \emph{DQN} as the proposed method, \emph{Aggressive} that injects reserve PRBs to approximately match the jammed PRBs using the same discrete action space, \emph{Idle} with no mitigation, and \emph{Random} that samples uniformly from $\mathcal{A}_{\mathrm{valid}}$.
In our experiments, we instantiate the step set as $\mathcal{D}=\{-9,-3,-1,0,1,3,9\}$ PRBs,
yielding $|\mathcal{A}|=49$ joint reserve-adjustment actions.
Metrics include URLLC latency, and reserved-PRB usage. 
We report three complementary metrics that capture service protection and reserve efficiency under jamming. 
(i) \emph{URLLC latency during jamming}, $L^{\mathrm{jam}}_{u}$, is computed by averaging the instantaneous URLLC queueing latency over all controller decision instants that fall inside jammer-ON windows within an episode, and then averaging across test episodes. 
(ii) \emph{Reserved-PRB usage during jamming}, $N^{\mathrm{jam}}_{\mathrm{used}}$, is the mean number of activated PRBs from the reserved pool during jammer-ON windows, again averaged over jammer-ON instants and then across episodes. 
(iii) \emph{Reserve-efficiency}, $\eta_L$, quantifies how effectively a policy converts reserve PRBs into URLLC protection and is defined as
$
\eta_L \triangleq \frac{1/L^{\mathrm{jam}}_{u}}{N^{\mathrm{jam}}_{\mathrm{used}}+\epsilon},
$
where $\epsilon$ avoids division by zero. Larger $\eta_L$ indicates lower jammer-ON latency achieved with fewer reserved PRBs.
Finally, to isolate \emph{jamming-induced inflation} from each policy’s inherent operating point, we also report the \emph{relative latency impact}
$
\Delta L^{\mathrm{jam}} \triangleq L^{\mathrm{jam}}_{u} - L^{\mathrm{base}}_{u},
$
where $L^{\mathrm{base}}_{u}$ is the same metric measured for that policy under no-jam conditions.

\begin{table}[htbp]
\caption{Simulation Parameters}
\label{tab:sim_params}
\centering
\begin{tabular}{|l|c|l|}
\hline
\textbf{Parameter} & \textbf{Symbol} & \textbf{Value} \\ \hline
Carrier frequency & $f_c$ & 3.5\,GHz (n78) \\ \hline
Bandwidth & $W$ & 100\,MHz \\ \hline
Cell / spawn radius & $R_{\text{cell}},R_{\text{spawn}}$ & 800\,m / 400\,m \\ \hline
Transmit power & $P_{tx}$ & 40\,dBm \\ \hline
REs per PRB & $\mathcal{N}_{\text{RE}}$ & 168 \\ \hline
Slot / step & $\mathcal{T}_{\text{slot}},T_{\text{step}}$ & $\approx1$\,ms , 1\,ms \\ \hline
Total PRBs / DL cap. & $N_{\text{total}},C_{\max}$ & 273 / 218 (80\% DL) \\ \hline
Base quotas & $q_e^{\min},q_u^{\min}$ & 140 / 60 PRBs \\ \hline
Reserved pool & $Z$ & 18 PRBs \\ \hline
UEs (eMBB/URLLC) & $|\mathcal{U}_e|,|\mathcal{U}_u|$ & 8 , 8 \\ \hline
Packet size (e/u) & $S_e,S_u$ & 1500 , 650 bytes \\ \hline
Jamming severity & $\mathcal{J}$ & $\{5,10,15,20\}$ PRBs \\ \hline
\end{tabular}
\end{table}

\subsection{Traffic load calibration and UE count}

Each UE generates one packet every 1 ms, yielding an arrival rate of 12 Mb/s per eMBB UE (1500 bytes per packet) and 5.2 Mb/s per URLLC UE (650 bytes per packet). Due to distance-dependent SNR and adaptive MCS selection, the achievable per-PRB rate spans 0.45–1.24 Mb/s/PRB. With the slice PRB quotas in Table~\ref{tab:sim_params}, the corresponding slice capacities are 63–174 Mb/s for eMBB and 27–74 Mb/s for URLLC. Comparing the aggregate arrival rate to these capacity bounds gives a feasible operating range of 5–14 UEs per slice. We set 8 UEs per slice to operate in a stressed-but-feasible regime, avoiding both trivial underload and persistent saturation.


\subsection{Experimental Results}


\begin{figure}[t]
\centering
\begin{subfigure}[t]{0.75\columnwidth}
  \centering
  \includegraphics[width=\linewidth]{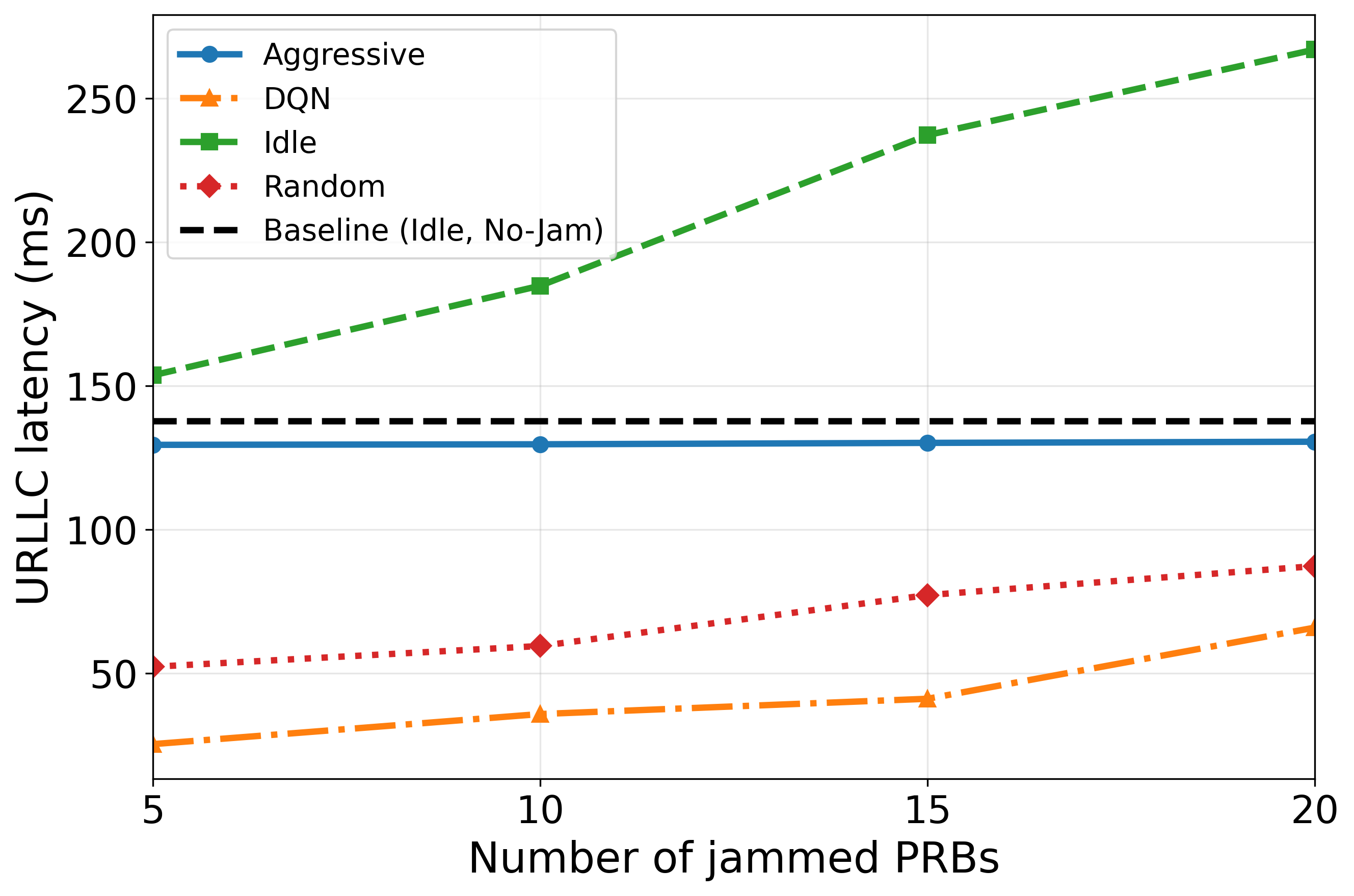}
  \caption{Mean URLLC latency.}
\end{subfigure}\hfill
\begin{subfigure}[t]{0.75\columnwidth}
  \centering
  \includegraphics[width=\linewidth]{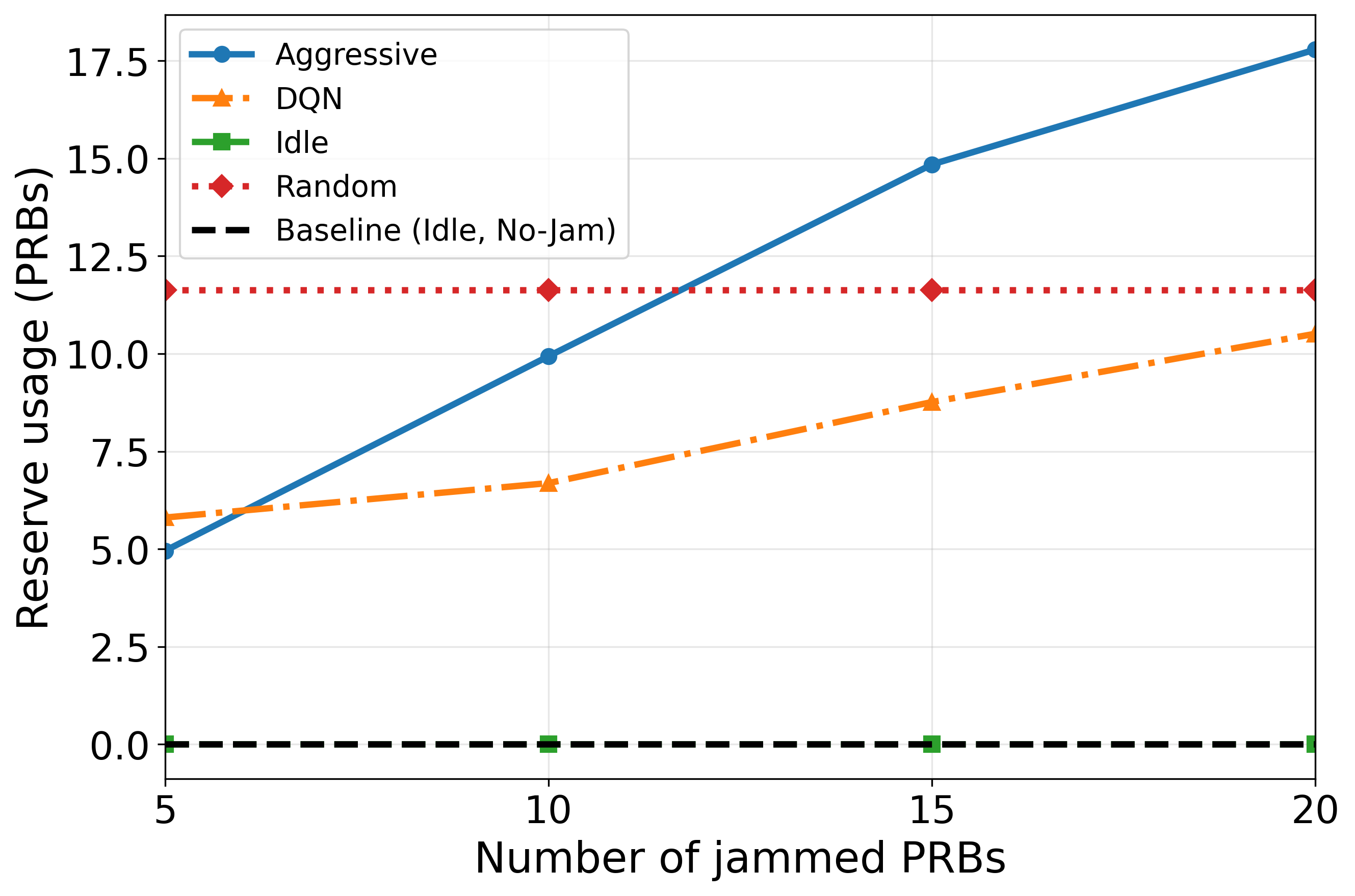}
  \caption{Mean reserved PRBs used.}
\end{subfigure}
\caption{Fixed-severity periodic on--off jamming sweep. Curves are jammer-ON averages over test episodes.\vspace{-0.1in}}
\label{fig:fixed_jamming_sweep}
\end{figure}

We evaluate four policies (DQN xApp, aggressive injection, idle, and random) under periodic on--off jamming that removes $J$
PRBs during jammer-ON.
Fig.~\ref{fig:fixed_jamming_sweep} evaluates periodic on--off jamming with fixed severity $\mathcal{J}$ and reports jammer-ON averages. In Fig.~\ref{fig:fixed_jamming_sweep}(a), Idle latency increases sharply with $\mathcal{J}$ because PRB removal amplifies queue buildup in the stressed regime, while the proposed DQN xApp keeps URLLC latency low by allocating reserve in response to emerging backlog and effective-capacity drops. Fig.~\ref{fig:fixed_jamming_sweep}(b) shows in the Aggressive approach, reserve injection rises sharply toward the pool limit as $\mathcal{J}$ grows, indicating over-provisioning and rapid budget consumption, whereas DQN scales reserve more selectively, preserving budget while sustaining URLLC protection.

\begin{figure}[t]
\centering
\includegraphics[width=0.37\textwidth]{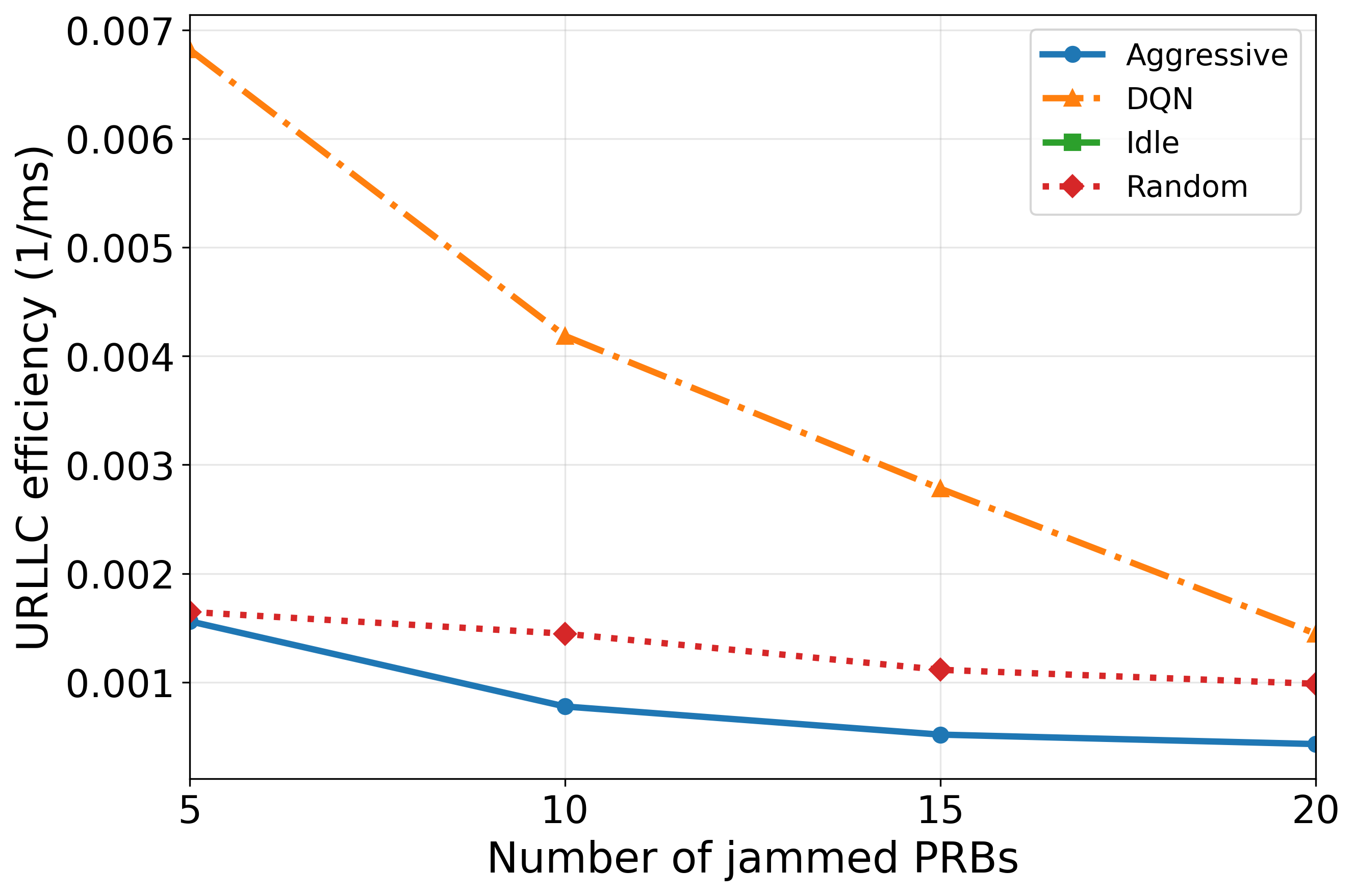}
\caption{Reserve-efficiency under fixed-severity jamming.}
\label{fig:reserve_efficiency}
\end{figure}


Fig.~\ref{fig:reserve_efficiency} reports reserve-efficiency $\eta_L$, i.e., URLLC protection achieved per reserved PRB used during jammer-ON. DQN attains the highest $\eta_L$ across severities, implying it spends reserve where it has the largest marginal impact on queue drain and latency reduction rather than simply matching jammed PRBs.

\begin{figure}[t]
\centering
\includegraphics[width=0.37\textwidth]{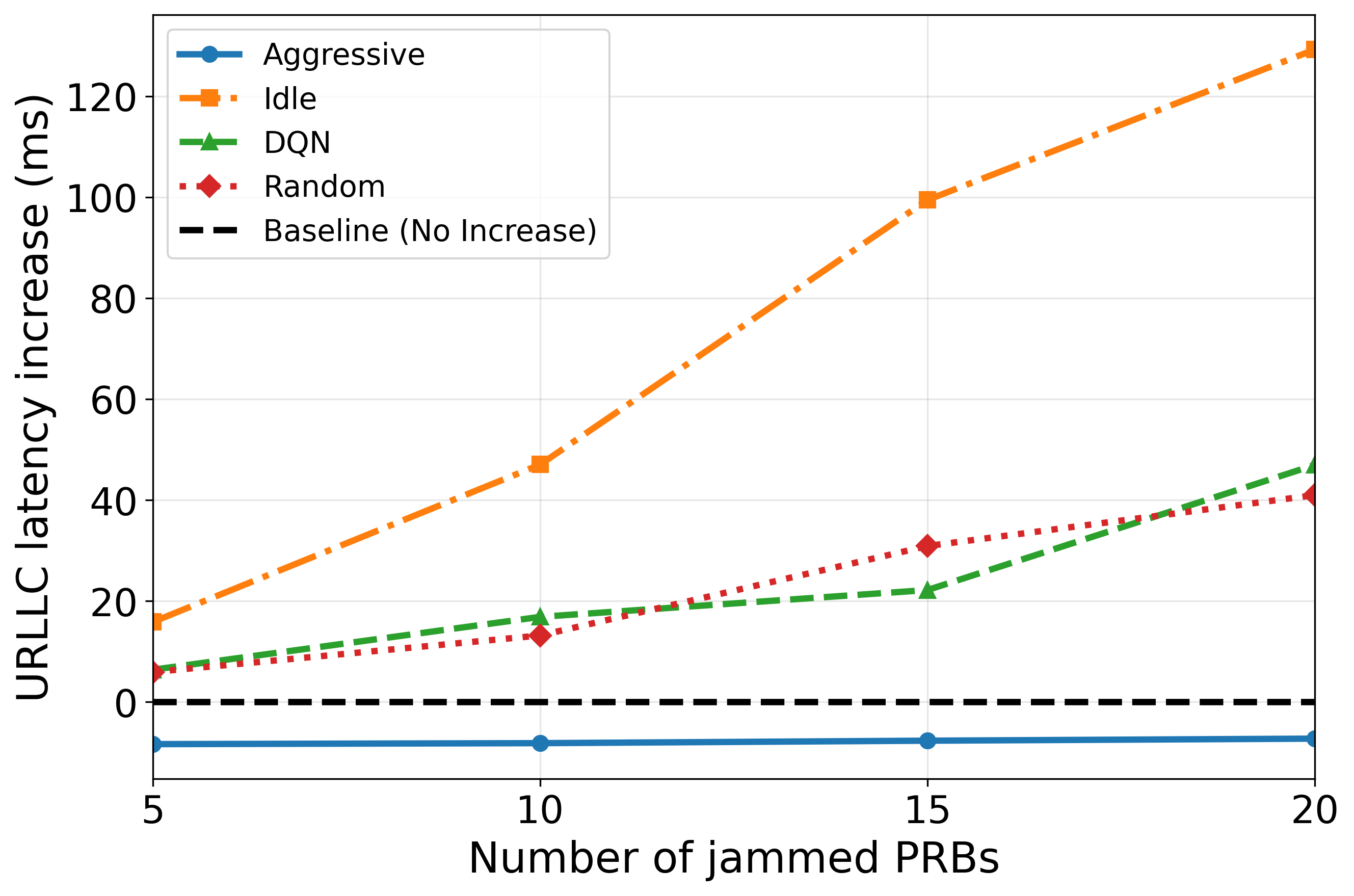}
\caption{Relative URLLC latency impact per-policy compared to no-jam baselines.}
\label{fig:relative_impact_fixed}
\end{figure}

Fig.~\ref{fig:relative_impact_fixed}
normalizes jammer-ON latency by each policy's no-jam baseline via $\Delta L^{\text{jam}}$, isolating attack-induced inflation from inherent operating points. DQN consistently reduces $\Delta L^{\text{jam}}$ versus Idle, while Aggressive degrades at higher severities as the finite reserve budget becomes binding; small negative $\Delta L^{\text{jam}}$ can occur when reserve PRBs persist briefly after jammer-OFF and accelerate backlog clearance.

\begin{figure}[t]
\centering
\includegraphics[width=0.45\textwidth]{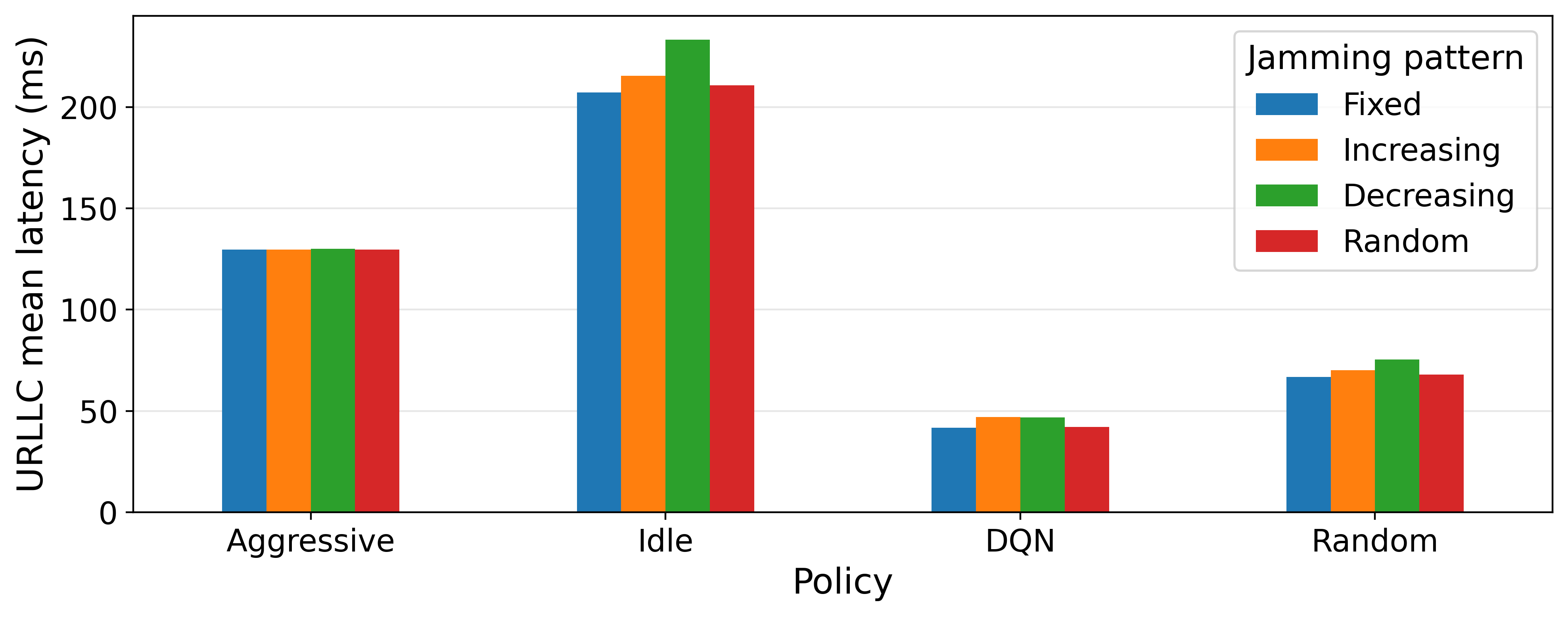}
\caption{URLLC latency during jammer-ON under time-varying severity profiles.}
\label{fig:pattern_lat}
\end{figure}


Fig.~\ref{fig:pattern_lat}
compares time-varying  severity profiles during jammer-ON. Idle performs worst due to unchecked backlog growth, whereas DQN maintains low jammer-ON latency across profiles, indicating robustness to within-interval severity drift and effective balancing of proactive backlog clearance with reactive reserve injection.

\section{Conclusion}
This paper addressed jamming-resilient PRB reservation in O-RAN network slicing by showing that capacity shocks caused by physical-layer jamming can induce persistent URLLC latency violations due to backlog-driven dynamics, particularly in the presence of low spectral-efficiency UEs. We formulated reserve activation as a time-coupled control problem with finite budget constraints and demonstrated that purely reactive mitigation is insufficient once congestion has accumulated. To address this, we designed a learning-based reserve control mechanism implemented as a Near-RT RIC xApp that combines proactive backlog clearance with reactive reserve allocation under jamming. Simulation results show that the proposed approach significantly reduces URLLC latency inflation while using reserved PRBs more efficiently than reactive and heuristic baselines, highlighting the importance of time-aware reserve control for resilience in software-driven O-RAN slicing systems.


\bibliographystyle{ieeetr} 
\bibliography{Resilient_allocation} 

@article{polese2023understanding,
  title={Understanding {O-RAN}: Architecture, interfaces, algorithms, security, and research challenges},
  author={Polese, Michele and Bonati, Leonardo and D’oro, Salvatore and Basagni, Stefano and Melodia, Tommaso},
  journal={IEEE Communications Surveys \& Tutorials},
  volume={25},
  number={2},
  pages={1376--1411},
  year={2023},
  publisher={IEEE}
}

@article{balasubramanian2021ric,
  title={{RIC}: A {RAN} intelligent controller platform for {AI}-enabled cellular networks},
  author={Balasubramanian, Bharath and Daniels, E Scott and Hiltunen, Matti and Jana, Rittwik and Joshi, Kaustubh and Sivaraj, Rajarajan and Tran, Tuyen X and Wang, Chengwei},
  journal={IEEE Internet Computing},
  volume={25},
  number={2},
  pages={7--17},
  year={2021},
  publisher={IEEE}
}

@article{luo2023field,
  title={Field trial of network slicing in {5G} and {PON-enabled} industrial networks},
  author={Luo, Yuanqiu and Jiang, Ming and Zhang, Dezhi and Effenberger, Frank},
  journal={IEEE Wireless Communications},
  volume={30},
  number={1},
  pages={78--85},
  year={2023},
  publisher={IEEE}
}

@inproceedings{moore2025demonstrating,
  title={Demonstrating Jamming Mitigation in {O-RAN} via {AI} enabled Intrusion Detection and Secure Slicing {xApps}},
  author={Moore, Joshua and Abdalla, Aly Sabri and Ueltschey, Charles and Marojevic, Vuk},
  booktitle={IEEE Military Communications Conference (MILCOM)},
    year= { Los Angeles, CA, USA, Oct. 2025},
}

@article{sharma2022mitigating,
  title={Mitigating jamming attack in 5G heterogeneous networks: A federated deep reinforcement learning approach},
  author={Sharma, Himanshu and Kumar, Neeraj and Tekchandani, Rajkumar},
  journal={IEEE Transactions on Vehicular Technology},
  volume={72},
  number={2},
  pages={2439--2452},
  year={2022},
  publisher={IEEE}
}

@techreport{3gpp_38214,
  author      = {3GPP},
  title       = {{NR; Physical layer procedures for data (Release 15)}},
  institution = {3rd Generation Partnership Project (3GPP)},
  year        = {2018},
  type        = {Technical Specification},
  number      = {TS 38.214 V15.3.0},
}

@inproceedings{kanervisto2020action,
  title={Action space shaping in deep reinforcement learning},
  author={Kanervisto, Anssi and Scheller, Christian and Hautam{\"a}ki, Ville},
  booktitle={IEEE conference on games (CoG)},
  year={ Osaka, Japan, Aug. 2020},
}

@techreport{ETSI_TR_138_901,
  title       = {{5G}; Study on channel model for frequencies from 0.5 to 100 {GHz} ({3GPP TR} 38.901 version 16.1.0 Release 16)},
  author      = {ETSI},
  institution = {European Telecommunications Standards Institute},
  number      = {TR 138 901 V16.1.0},
  month       = {Nov.},
  year        = {2020}
}

@article{balapuwaduge2018dynamic,
  title={Dynamic Spectrum Reservation for {CR} Networks in the Presence of Channel Failures: Channel Allocation and Reliability Analysis},
  author={Balapuwaduge, Indika AM and Li, Frank and Pla, Vicent},
  journal={IEEE Transactions on Wireless Communications},
  volume={17},
  number={2},
  pages={882--898},
  year={2018},
  publisher={Institute of Electrical and Electronics Engineers}
}

@article{el2020centralized,
  title={Centralized dynamic channel reservation mechanism via {SDN} for {CR} networks spectrum allocation},
  author={El Azaly, Nehal M and Badran, Ehab F and Kheirallah, Hassan Nadir and Farag, Hania H},
  journal={IEEE Access},
  volume={8},
  pages={192493--192505},
  year={2020},
  publisher={IEEE}
}

@article{abbas2020spectrum,
  title={Spectrum efficiency in {CRNs} using hybrid dynamic channel reservation and enhanced dynamic spectrum access},
  author={Abbas, Ghulam and Abbas, Ziaul Haq and Baker, Thar and Waqas, Muhammad and others},
  journal={Ad Hoc Networks},
  volume={107},
  pages={102246},
  year={2020},
  publisher={Elsevier}
}

@article{khan2021enhanced,
  title={An Enhanced Spectrum Reservation Framework for Heterogeneous Users in {CR-Enabled IoT} Networks},
  author={Khan, Abd Ullah and Tanveer, Muhammad and Khan, Wali Ullah and Nebhen, Jamel and Li, Xingwang and Zeng, Ming and Dobre, Octavia A},
  journal={IEEE Wireless Communications Letters},
  year={2021},
  publisher={IEEE Communications Society, Piscataway, United States-New Jersey}
}

@article{luis2016novel,
  title={A novel reservation-based {MAC} scheme for distributed cognitive radio networks},
  author={Lu{\'\i}s, Miguel and Oliveira, Rodolfo and Dinis, Rui and Bernardo, Luis},
  journal={IEEE Transactions on Vehicular Technology},
  volume={66},
  number={5},
  pages={4327--4340},
  year={2016},
  publisher={IEEE}
}

@article{houda_federated_2024,
	title = {Federated {Deep} {Reinforcement} {Learning} for {Efficient} {Jamming} {Attack} {Mitigation} in {O}-{RAN}},
	volume = {73},
	issn = {1939-9359},
	url = {https://ieeexplore.ieee.org/document/10416344/},
	doi = {10.1109/TVT.2024.3359998},
	number = {7},
	urldate = {2026-01-09},
	journal = {IEEE Transactions on Vehicular Technology},
	author = {Houda, Zakaria Abou El and Moudoud, Hajar and Brik, Bouziane},
	month = jul,
	year = {2024},
	pages = {9334--9343},
}

@article{asemian_anti-jamming_2025,
	title = {Anti-{Jamming} {Task} {Scheduling} in {MEC}-{O}-{RAN} with {Hierarchical} {DRL} and {Transformer}-{Based} {Control}},
	issn = {2327-4662},
	url = {https://ieeexplore.ieee.org/document/11278191/},
	doi = {10.1109/JIOT.2025.3640520},
	urldate = {2026-01-09},
	journal = {IEEE Internet of Things Journal},
	author = {Asemian, Ghazal and Amini, Mohammadreza and Kantarci, Burak},
	year = {2025},
	pages = {1--1},
}

@article{pirayesh_jamming_2022,
	title = {Jamming {Attacks} and {Anti}-{Jamming} {Strategies} in {Wireless} {Networks}: {A} {Comprehensive} {Survey}},
	volume = {24},
	issn = {1553-877X},
	shorttitle = {Jamming {Attacks} and {Anti}-{Jamming} {Strategies} in {Wireless} {Networks}},
	url = {https://ieeexplore.ieee.org/document/9733393/},
	doi = {10.1109/COMST.2022.3159185},
	number = {2},
	urldate = {2026-01-09},
	journal = {IEEE Communications Surveys \& Tutorials},
	author = {Pirayesh, Hossein and Zeng, Huacheng},
	year = {2022},
	pages = {767--809},
}

@article{thanh_anti-jamming_2022,
	title = {Anti-{Jamming} {RIS} {Communications} {Using} {DQN}-{Based} {Algorithm}},
	volume = {10},
	issn = {2169-3536},
	url = {https://ieeexplore.ieee.org/document/9732955/},
	doi = {10.1109/ACCESS.2022.3158751},
	urldate = {2026-01-09},
	journal = {IEEE Access},
	author = {Thanh, Pham Duy and Giang, Hoang Thi Huong and Hong, Ic-Pyo},
	year = {2022},
	pages = {28422--28433},
}

@misc{airansim,
  title        = {{AI-RAN} Simulator},
  howpublished = {\url{https://github.com/ntutangyun/ai-ran-sim}},
  note         = {Accessed: 2025-11-24}
}
\end{document}